\documentclass[epj,referee]{svjour}

\usepackage{graphicx}

\begin{document}

\newcommand{\EQ}{Eq.~}
\newcommand{\EQS}{Eqs.~}
\newcommand{\FIG}{Fig.~}
\newcommand{\FIGS}{Figs.~}
\newcommand{\TAB}{Tab.~}
\newcommand{\SEC}{Sec.~}
\newcommand{\SECS}{Secs.~}

\title{Subcritical behavior in the alternating
supercritical Domany-Kinzel dynamics}
\author{Naoki Masuda\inst{1,2} \and Norio Konno\inst{3}}
\institute{Laboratory for Mathematical Neuroscience,
RIKEN Brain Science Institute, 2-1, Hirosawa, Wako, Saitama, 351-0198,
Japan \and
Aihara Complexity Modelling Project, ERATO, JST, 3-23-5, Uehara,
Shibuya, Tokyo, 151-0064, Japan
\and
Faculty of Engineering,
Yokohama National University,
79-5, Tokiwadai, Hodogaya, Yokohama, 240-8501, Japan
\email{masuda@brain.riken.jp}
}
%norio@mathlab.sci.ynu.ac.jp
% \date{}
\date{Received: date / Revised version: date}

\abstract{
Cellular automata are widely used to model real-world
dynamics.
We show using the Domany-Kinzel probabilistic
cellular automata that alternating
two supercritical dynamics can result in 
subcritical dynamics in which the population dies out. The
analysis of the original and reduced
models reveals generality of this paradoxical behavior, which suggests that
autonomous or man-made periodic
or random environmental changes can cause extinction in otherwise safe
population dynamics. Our model also
realizes another scenario for the Parrondo's paradox to occur, namely,
spatial extensions.
}
\maketitle
\section{Introduction}\label{sec:introduction}

Ecological and sociological dynamics are often described by systems of
locally interacting agents. Cellular automata are broadly used for
modeling such dynamics to characterize, for example, 
survival probability, percolation, and critical
phenomena, which are relevant to real situations
\cite{Wolframbook}.  Among the class of probabilistic
cellular automata is the Domany-Kinzel (DK) model, which is a two
parameter family of Markov processes on a one-dimensional lattice with
discrete time \cite{Domany,Kinzel}.  In this paper, we report a
counterintuitive phenomenon of the DK model: particles eventually die
out when two supercritical DK dynamics
alternate with some appropriate orders.
This behavior is robust against parameter
changes. We also analyze the reduced dynamics such as
the pair approximation and a canonical model to guarantee that
this phenomenon is preserved in much simpler models. As a generalization,
dynamic environmental changes can extinguish a
population even if the snapshot dynamics
is supercritical at any given moment. These
alternating DK dynamics also realize a new scenario for the Parrondo's
paradox \cite{Harmer,Parrondo,Harmer02} to occur,
that is, introduction of the space.

\section{DK model}

\begin{figure}
\begin{center}
\includegraphics[height=2.25in,width=2.75in]{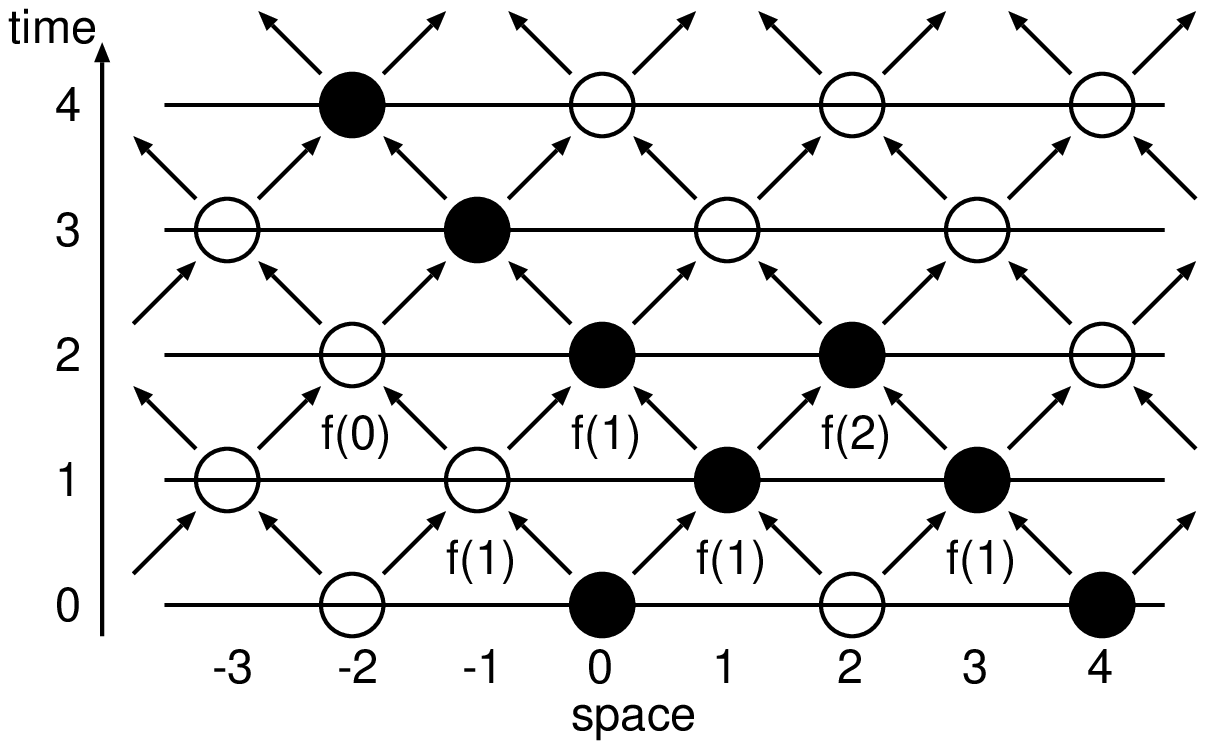}
\caption{}
\label{fig:dk_pic}
\end{center}
\end{figure}

In the DK model \cite{Domany,Kinzel}, each site either accompanies a
particle (denoted by $\bullet$) or is empty (denoted by $\circ$) at
any instant.  The space can be identified with a subset of the set of
integers ${\bf Z}$, and let $\xi_n \subset {\bf Z}$ be the set of the
sites that have particles at discrete time $n\in {\bf Z}_+$ $=
\{0,1,2, \ldots \}$. The stochastic evolution rule at each site $x\in {\bf
Z}$ is independently described by $P( x \in \xi_{n+1} | \xi_n ) = f (
|\xi_n \cap \{x-1, x+1\}|)$ where $f(0)=0$, $f(1)=p_1$, $f(2)=p_2$,
and $(p_{1}, p_{2}) \in [0,1]^{2}$. In other words, the probability
that a particle emerges is determined by the number of the
particles in the nearest neighborhood in the previous time, as shown
in \FIG\ref{fig:dk_pic}.  Each realization of the spatiotemporal
process is expressed in the form of a configuration $\xi \in
\{0,1\}^{\bf S}=X$ with ${\bf S}= \{ s=(x,n)\in {\bf Z} \times {\bf
Z}_{+} : x+n= \hbox{even} \}$. The region of the supercritical
parameter sets ($p_1, p_2$) for which particles survive for infinite
time with positive probability can be numerically obtained, and it
occupies an upper-right area in the $p_1$-$p_2$ space
\cite{Domany,Kinzel,Durrettbook}.  The DK model is equivalent to the
directed bond percolation model on a square lattice when $(p_1, p_2)$
$=$ $(p, 2p - p^2)$ and to the directed site percolation model when
$p_1 = p_2 = p$ \cite{Domany,Kinzel,Durrettbook}.  Another special
case is Wolfram's rule 90 deterministic cellular automaton
\cite{Wolframbook} which is realized with $(p_{1}, p_{2}) = (1,0)$. The
simplicity of the DK model enables us to investigate interesting
properties from the viewpoint of statistical physics and applications,
such as quasistationary particle density
\cite{Martins,Gutowitz,Tome,Harada,Atman02},
critical phenomena and phase transitions
\cite{Domany,Kinzel,Martins,Gutowitz,Tome,Harada,Atman02,Bagnoli,Atman03},
survival probabilities \cite{Katori}, and duality
\cite{Konno02a,Konno02b,Katori04}.

Let us denote by $P_n(\cdot)$ the probability that an event occurs at
time $n$. Here an event means a state of consecutive sites, or a
sequence of $\bullet$ and $\circ$. For clarity, we often plot
trajectories in the two-dimensional space spanned by the
order parameters defined with $a_2(n)\equiv P_n(\bullet\bullet)$ and
$a_1(n)\equiv P_n(\bullet\circ) + P_n(\circ\bullet)$. 
With
$\protect{a_0(n) \equiv P_n(\circ\circ)}$, it follows that
$\protect{a_1(n) \ge 0}$,
$a_2(n)\ge 0$, and $a_1(n)+a_2(n)$ $=$ $1-a_0(n)\le 1$. The origin $(a_1,
a_2)$ $=$ $(0, 0)$ is an absorbing fixed point corresponding to the
population death.  In the
following numerical simulations, the lattice size is $10000$, and the
periodic boundary conditions are assumed.

With some initial conditions, trajectories of the DK model are shown
in \FIG\ref{fig:dk}(a) for ($p_1, p_2$) $=$ (0.52, 1) (thin lines) and
for (0.76, 0.76) (thick lines). The DK dynamics corresponding to these
parameter sets are termed dynamics $A$ and dynamics $B$, respectively.
When $p_2=1$, particles emerge or die only at kinks where $\bullet$
and $\circ$ face each other.  In this case, the dynamics of kinks are
identical to the coalescing random walk, and the entire space is
eventually occupied by particles with a positive probability if and
only if $p_1>0.5$ \cite{Durrettbook}. Therefore, dynamics $A$ is
supercritical.  On the other hand, the DK model with $p_1=p_2$ is
equivalent to the directed site percolation. Restricted onto this line,
$p_1 = p_2 = 0.75$ is a mathematically rigorous upper bound for the
subcritical regime \cite{Liggett95}, whereas the critical value
is numerically estimated to be about $p_1 = p_2 = 0.7055$
\cite{Onody,Jensen}. Because of the attractiveness
($p_1\le p_2$), the natural intuition that more particles are likely to
survive with larger $p_1$ and $p_2$ actually holds
\cite{Durrettbook,Katori}. Therefore, dynamics $B$ is also
supercritical.  Accordingly, trajectories of dynamics $A$ converge to
the all $\bullet$ state, and those of $B$ converge to the stochastic
stable fixed point $(a_1, a_2) \cong (0.39, 0.42)$.

\section{Population Death in Alternating DK Dynamics}

\begin{figure}
\begin{center}
\includegraphics[height=2.75in,width=2.75in]{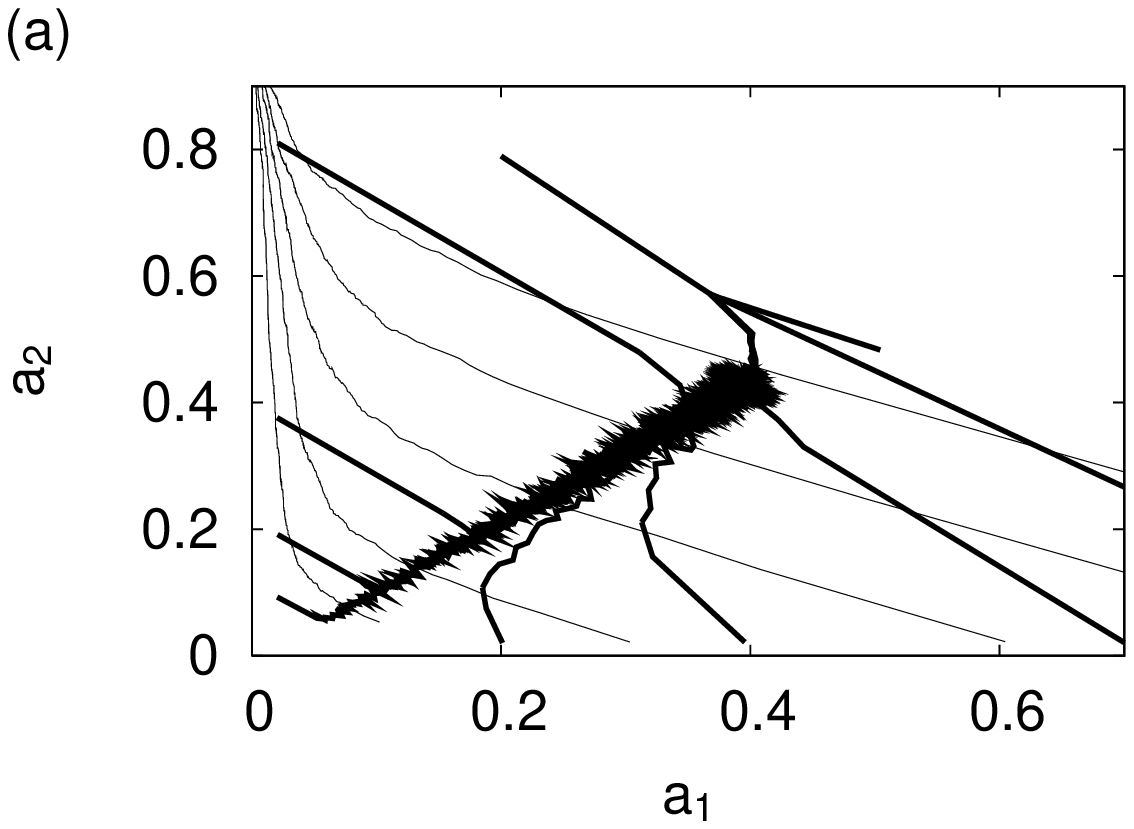}
\includegraphics[height=2.75in,width=2.75in]{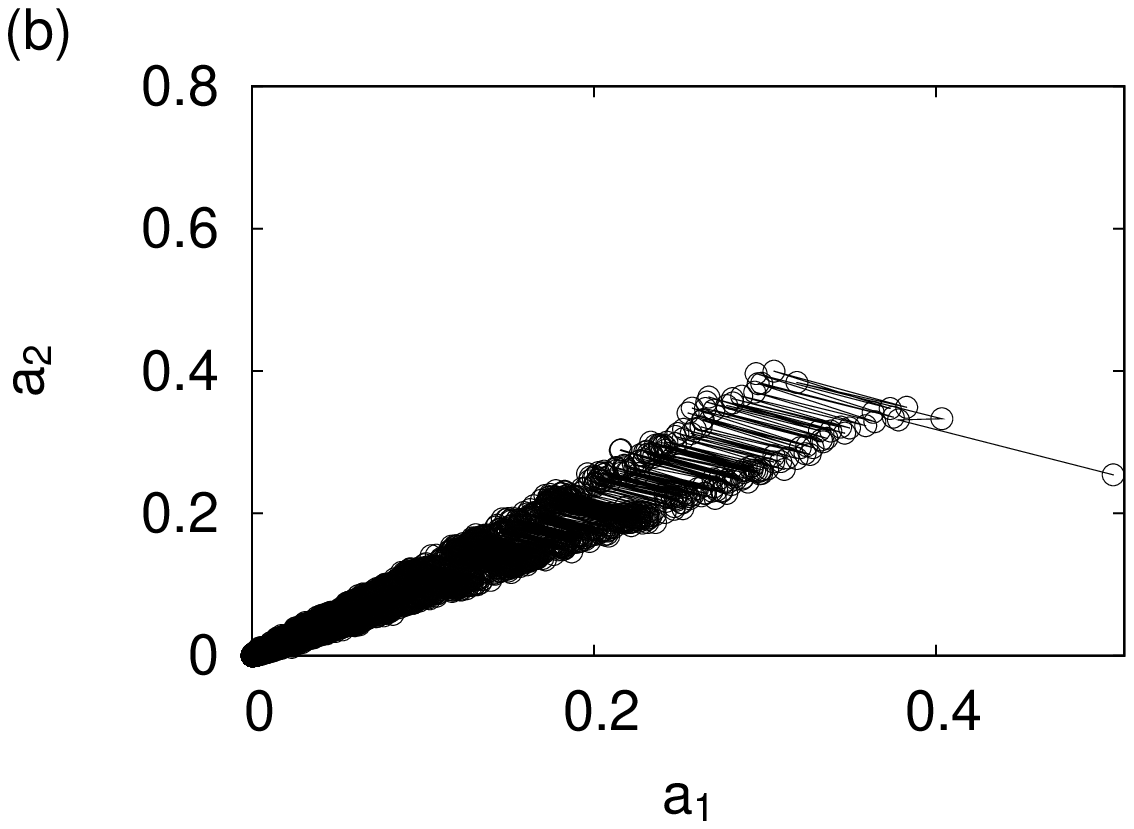}
\includegraphics[height=2.75in,width=2.75in]{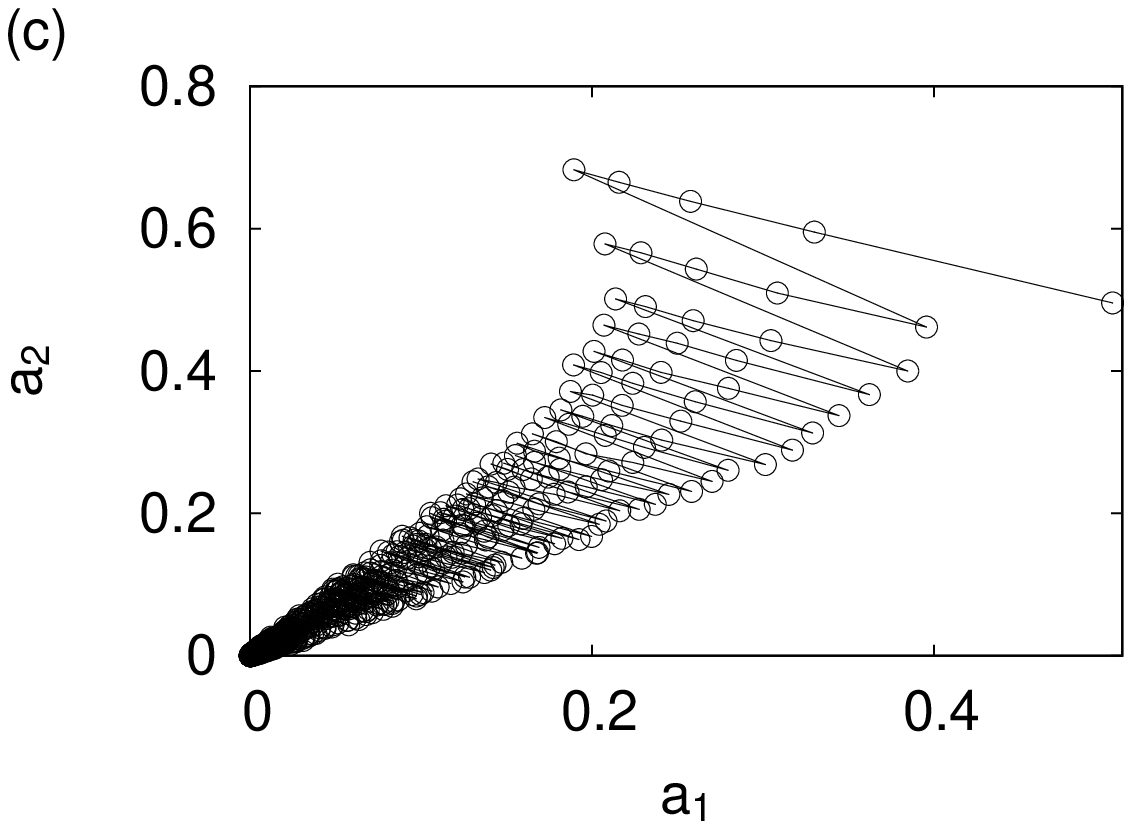}
\includegraphics[height=2.75in,width=2.75in]{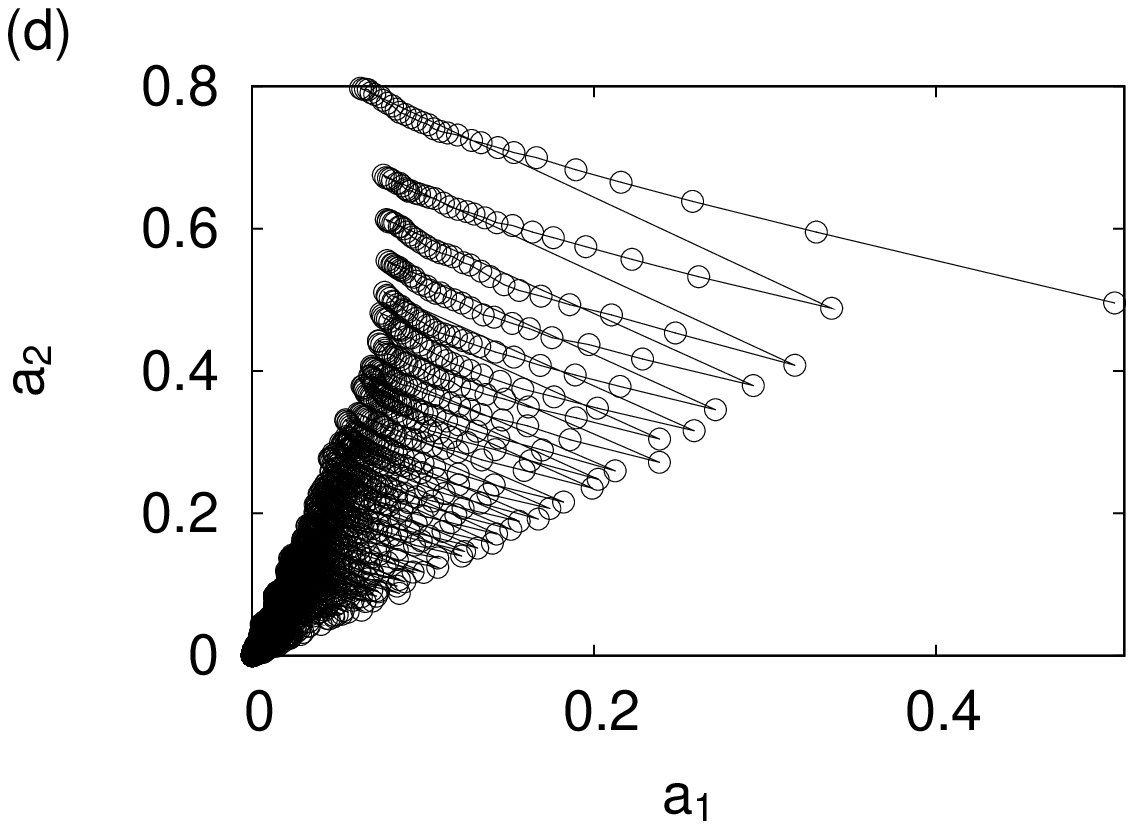}
\caption{}
\label{fig:dk}
\end{center}
\end{figure}

Next, we alternatively apply $A$ and $B$. A typical trajectory
is shown in \FIG\ref{fig:dk}(b) with the Bernoulli initial distribution with
density 0.5, which
yields $(a_1(0), a_2(0)) = (0.5, 0.25)$. Surprisingly, particles eventually
die out. This behavior is not sensitive to the choice of initial
conditions.  It also persists against changes in $p_1$ or $p_2$ as far
as the individual dynamics are not extremely supercritical and the
stable stochastic fixed points for the two systems are separated
enough. Especially, extensive
numerical simulations suggest that this population
death is enhanced when one of the component dynamics is nonattractive, or
$p_1>p_2$.

\begin{figure}
\begin{center}
\includegraphics[height=2.75in,width=2.75in]{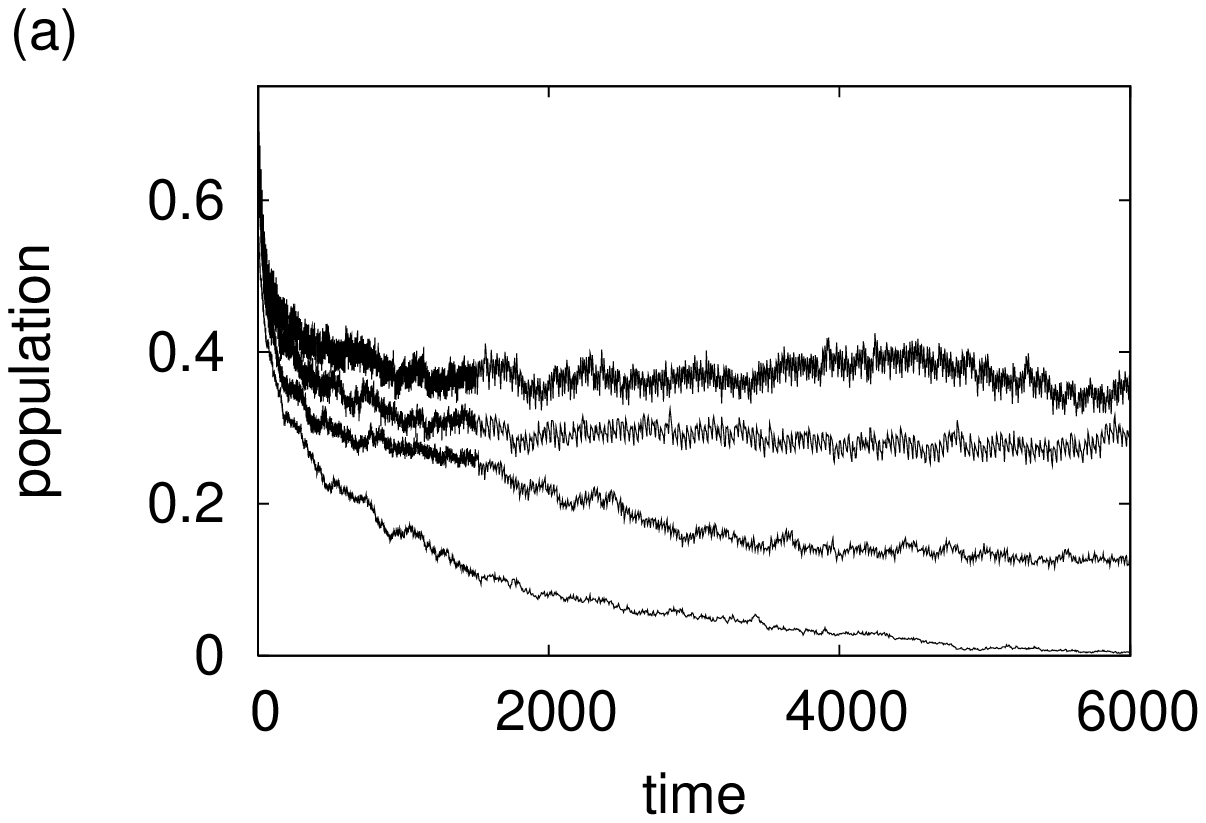}
\includegraphics[height=2.75in,width=2.75in]{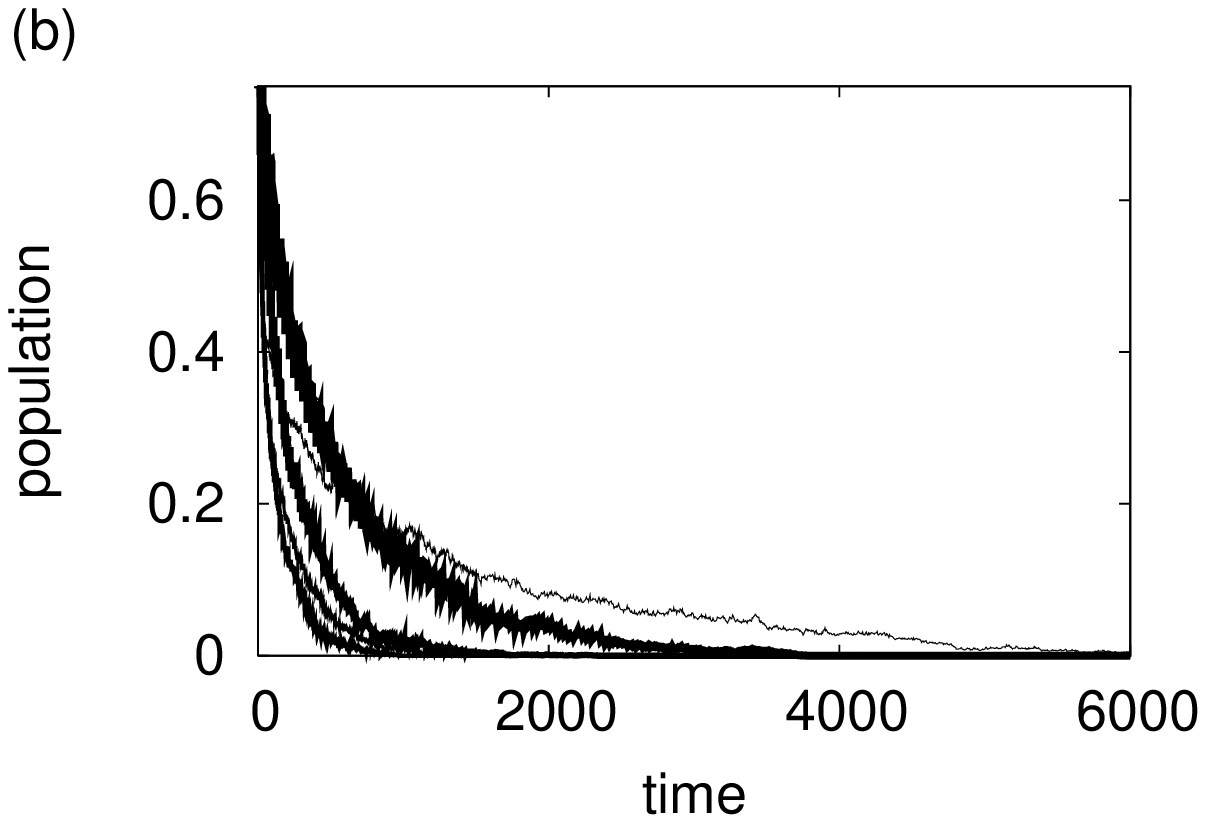}
\includegraphics[height=2.75in,width=2.75in]{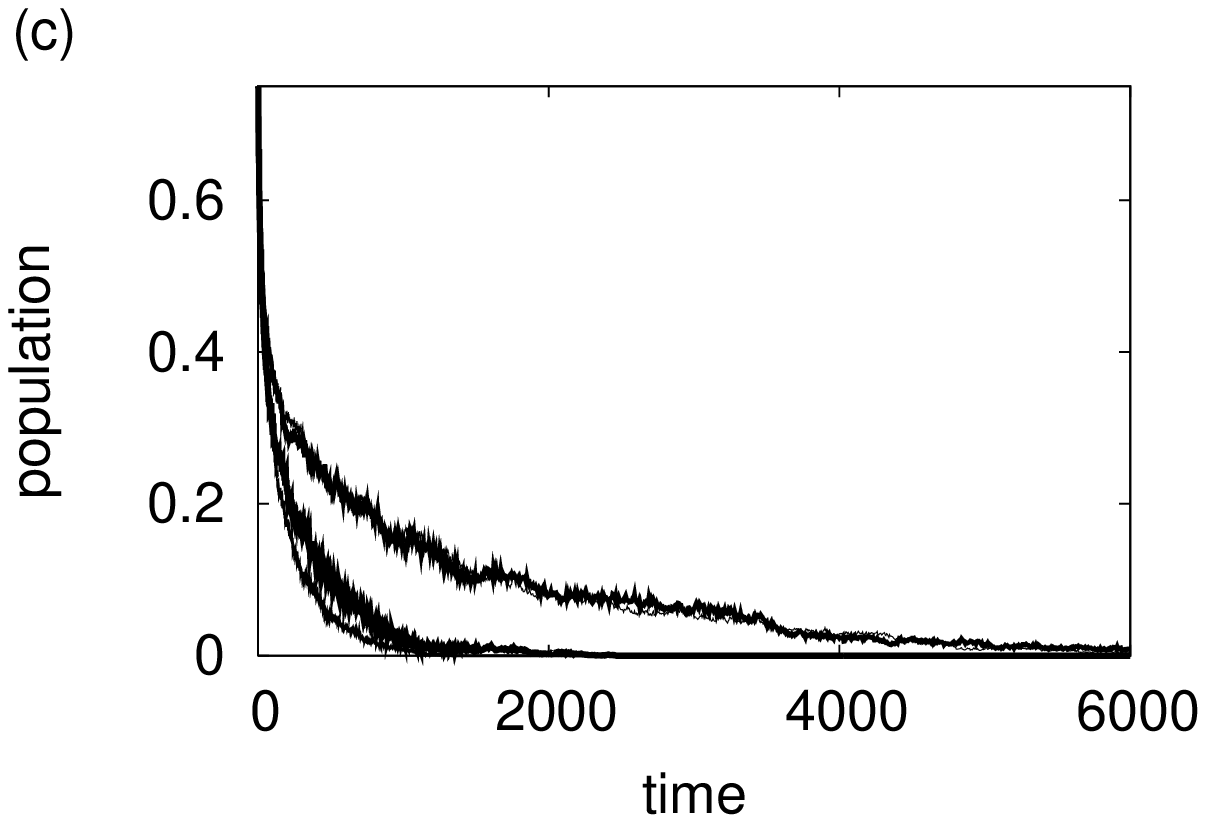}
\caption{}
\label{fig:sdk}
\end{center}
\end{figure}

An important cause for the population death is how the trajectories of
dynamics $A$ and those of dynamics $B$ cross. As shown in
\FIG\ref{fig:dk}(a), if a state in the $a_1$-$a_2$ space evolves along
a trajectory of dynamics $A$, in terms of dynamics $B$,
the state gradually slides down to trajectories associated with
initial conditions with fewer particles. In other words, from the
viewpoint of dynamics $A$ (resp. $B$), the population once decreases
under dynamics $B$ (resp. $A$) before it revives and reaches the
nontrivial stable fixed point.
Therefore, by switching the dynamics between $A$
to $B$ before the population effectively starts to grow, the number of
the particles
gradually decrease to zero. Survival results if $A$ or $B$ is
applied long enough before switching to the other.  To demonstrate
this, we confine ourselves to the cases in which a block of $k$ $A$'s
and $k$ $B$'s are alternatively applied, which we denote by $A{}^k
B{}^k$.  As shown in \FIG\ref{fig:sdk}(a), the population is more
likely to survive as $k$ increases.

The population death by alternation is an example of the Parrondo's
game in which a combination of two losing (winning) stochastic games
can counterintuitively end up with a winning (losing) game
\cite{Harmer,Parrondo,Harmer02,Amengual}.  In this context, the
results in \FIG\ref{fig:sdk}(a) agree with those for the
original Parrondo's game in which the paradoxical effect becomes small
as $k$ is raised \cite{Harmer02}.  A more general concern is how the
arrangement of $A$ and $B$ affects the upshot.  Since it appears
quite difficult to derive the optimal ordering of $A$ and $B$ among
all the possible sequences \cite{Harmer02,Amengual}, we only deal with
some representative cases.

The population dynamics when a chain of $A$ is periodically punctuated
by just one $B$, which is denoted by $A^k B$, are shown in
\FIG\ref{fig:sdk}(b). This figure together with additional numerical
simulations suggests that the paradoxical effect is most manifested,
or the population dies out most rapidly, with $k=4$. This is
presumably because dynamics $B$ correspond to the critical line of the
attractiveness ($p_1=p_2$).  For this reason, in an upper-left region
in the $a_1$-$a_2$ space, an application of the near-nonattractive $B$
kills more particles when there exist more particles. This view is
supported by \FIGS\ref{fig:dk}(b) and \ref{fig:dk}(c) in which we
compare the dynamics with $AB$ and those with $A^4B$. Then, the
convergence to $(a_1, a_2)=(0,0)$ is accelerated by a larger $k$ in a
small $k$ regime.  However, with a much larger $k$, the number of
particles changes little for most of the time (\FIG\ref{fig:dk}(d)).
In this regime, the population death is slowed down as $k$ increases.

For sequences in the form $A B^k$, the parity effect is manifested.  As
shown in \FIG\ref{fig:sdk}(c), the population death is faster when $k$
is even.  This is again because dynamics $B$ is nearly nonattractive.
As is prominent in nonattractive DK dynamics, the motion in the
$a_1$-$a_2$ space under dynamics $B$ is somewhat sensitive to the
current state. More specifically, simple repetition of $B$ yields a
damped oscillation in the early stage. Therefore, if the initial state
is located in a upper-left region, the number of particles drops more
when $B$ is repeated even times before being interrupted by one $A$.
 
\begin{figure}
\begin{center}
\includegraphics[height=2.75in,width=2.75in]{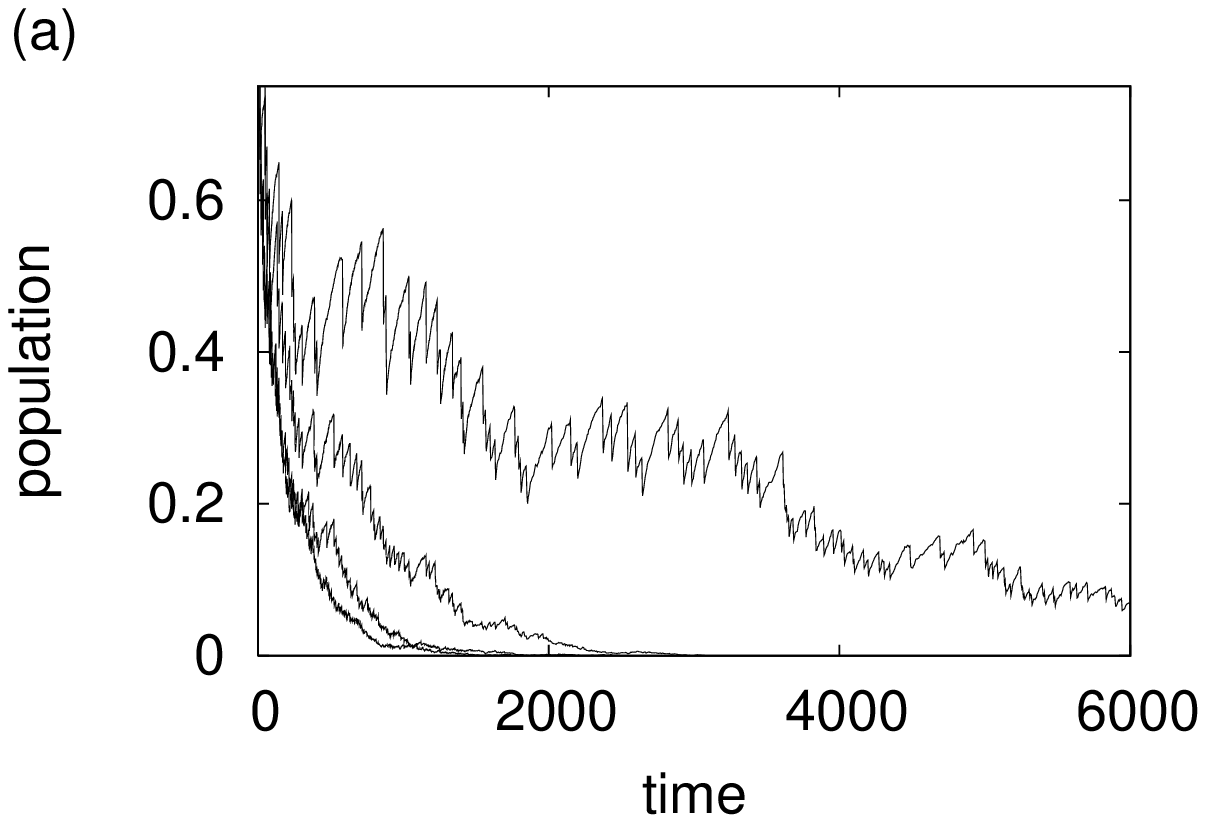}
\includegraphics[height=2.75in,width=2.75in]{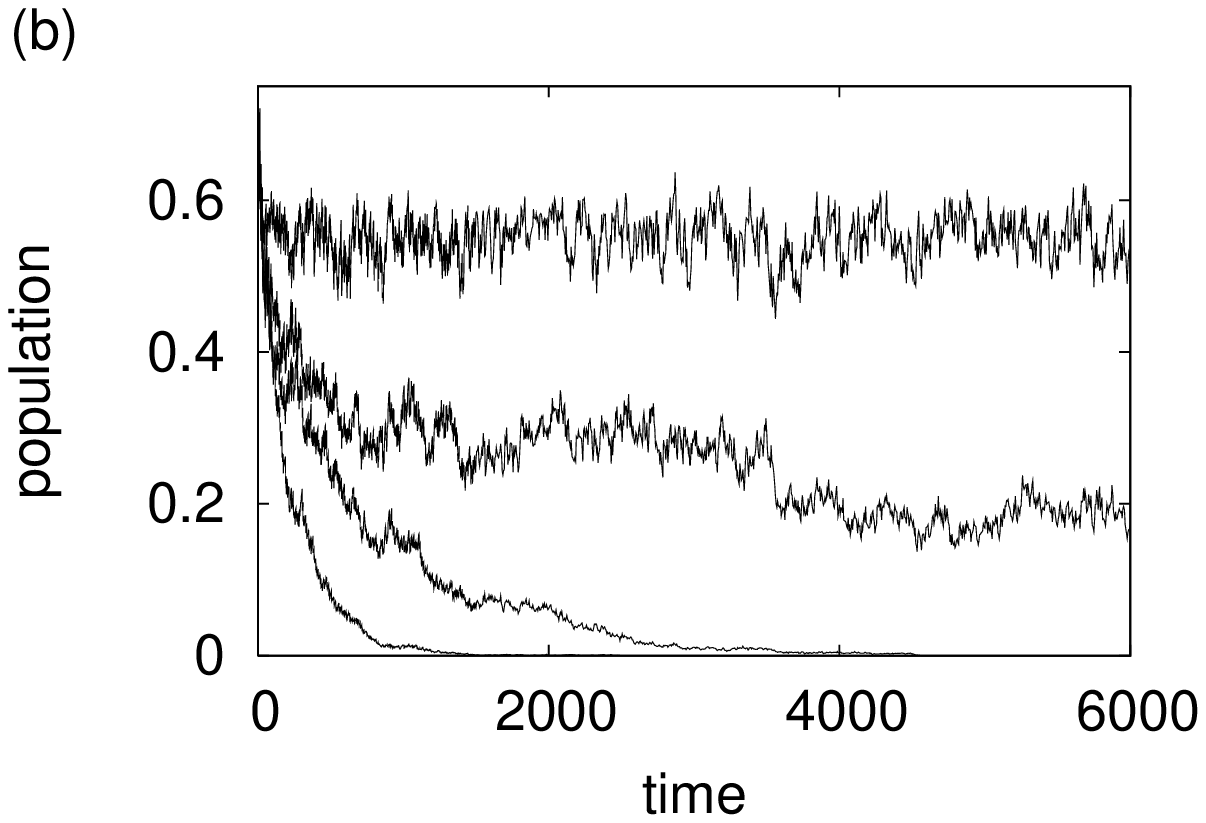}
\caption{}
\label{fig:sdkr}
\end{center}
\end{figure}

The random arrangement of $A$ and $B$ is also of interest
\cite{Harmer,Harmer02,Amengual} because real environments
can be random rather than perfectly periodic. To mimic simple random
environments, we choose $A$ and $B$
independently at each time step with probability $r$ and $1-r$,
respectively. Obviously, the population death does not occur with $r=0$ or with
$r=1$, which prescribes the sequence purely of $B$ and that of $A$,
respectively.  Figure~\ref{fig:sdkr} and the extensive parameter
search reveal that the paradoxical effect is maximized when $r\cong
0.2$. This value coincides with the optimal 
mixing ratio
for the family of deterministic sequences investigated above, namely, $A^4B$.

\begin{figure}[h]
\begin{center}
\includegraphics[height=2.75in,width=2.75in]{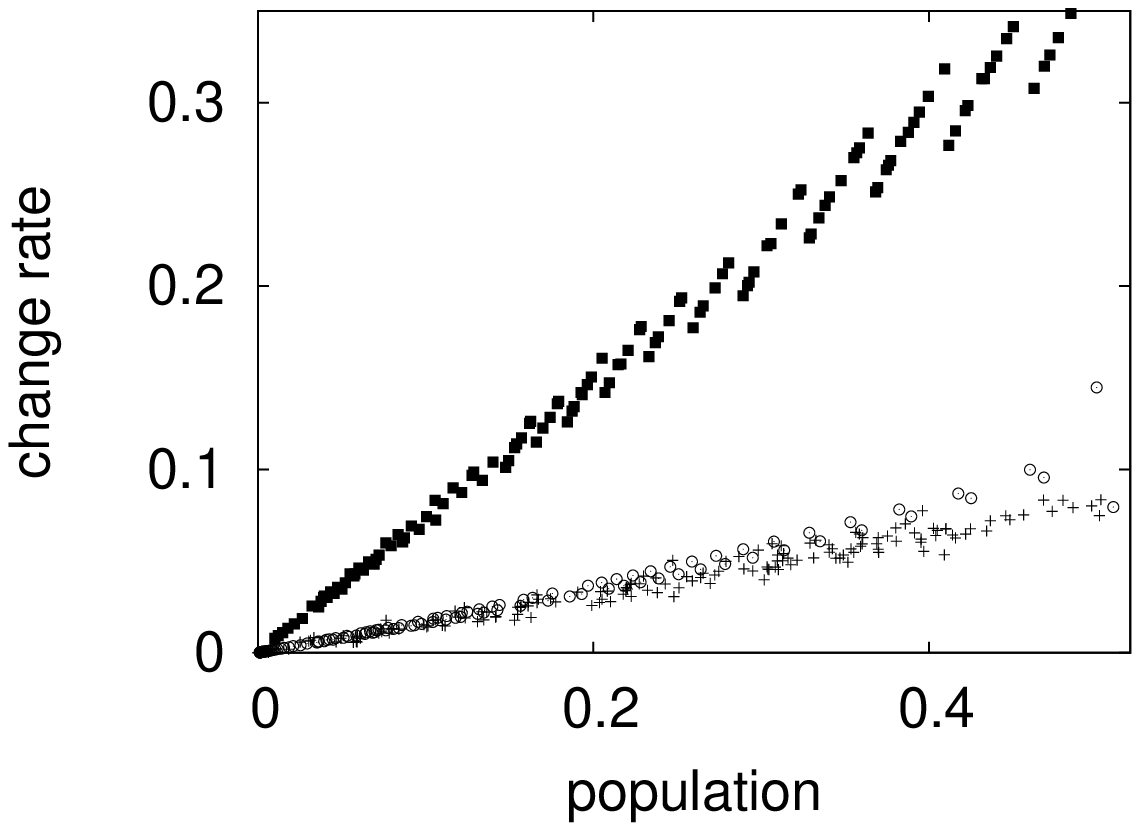}
\caption{}
\label{fig:stat}
\end{center}
\end{figure}

It is also essential for the paradox
that population change rates are proportional to the population size
as shown in \FIG\ref{fig:stat} (crosses).  Owing to this property, the
size of the population exponentially shrinks to a very small level
(\FIGS\ref{fig:dk}, \ref{fig:sdk}, \ref{fig:sdkr}).
Then, particles become 
extinct in finite time
because of stochasticity and the absolute stability of the fixed
point $(a_1, a_2)$ $=$ $(0,0)$.
If change rates are too high even for
minute population mass, a trajectory that happens to have approached the
origin more likely escapes the vicinity of the origin to avoid
the population death.

\section{Pair Approximation}

\begin{figure}[h]
\begin{center}
\includegraphics[height=2.75in,width=2.75in]{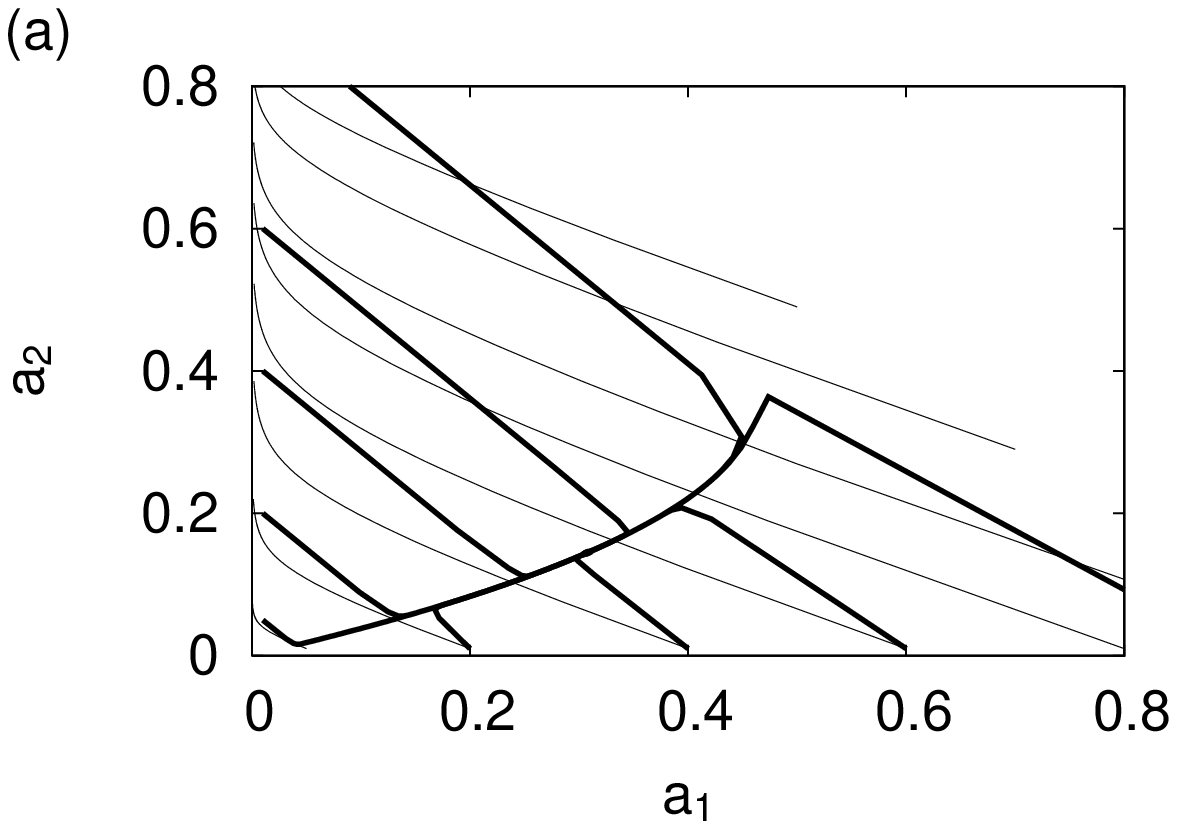}
\includegraphics[height=2.75in,width=2.75in]{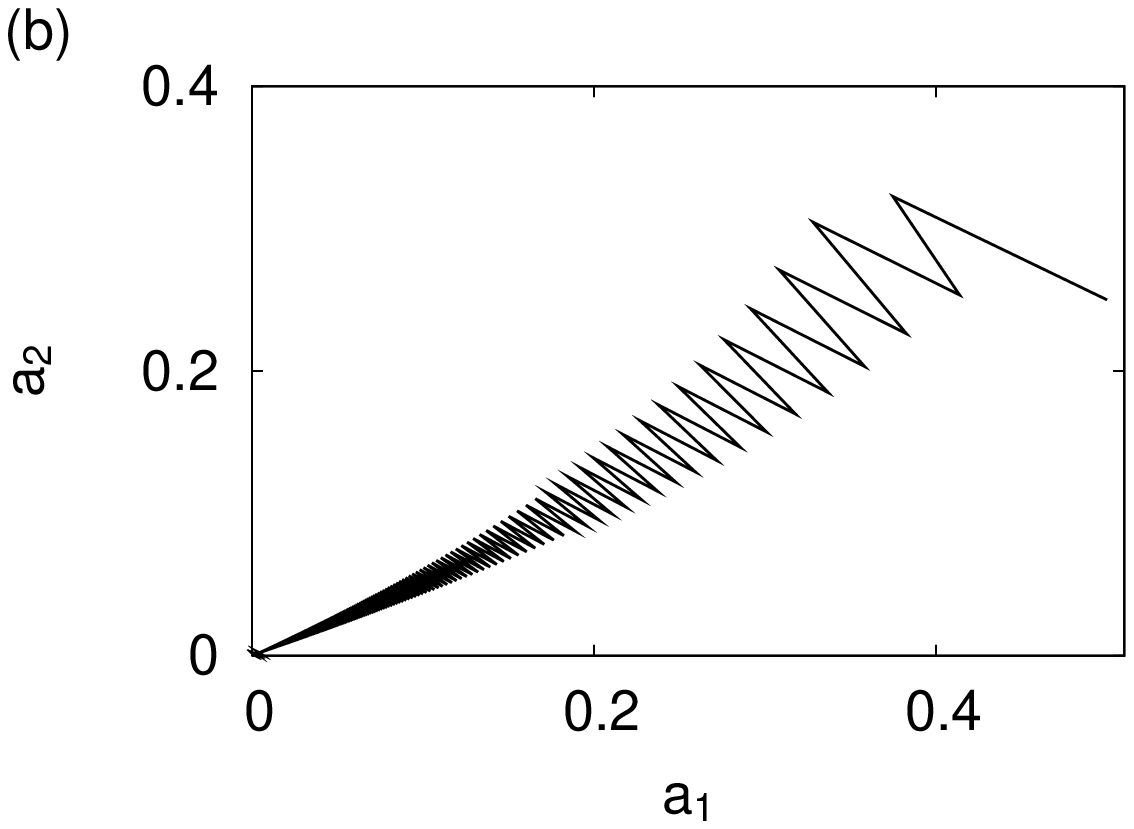}
\caption{}
\label{fig:pair}
\end{center}
\end{figure}

To take a closer look at the paradox,
we analyze the deterministic
dynamics derived by the pair approximation of the DK model, which
we call the PA dynamics \cite{Gutowitz,Tome,Harada,Atman02}.
In the pair approximation, any events at two
sites separated by a distance more than one are supposed to be
independent of each other.
For example, $P_n(\bullet | \circ\bullet)$ $=$ 
$P_n(\bullet \circ\bullet) / P_n(\circ\bullet)$
is approximated by
$P_n(\bullet | \circ)$ 
$=$ $P_n(\bullet\circ) / P_n(\circ)$,
where $P_n(\cdot |\cdot)$ denotes the conditional
probability. Accordingly, probabilities of any events involving three
or more consecutive sites are decomposed into 
one- or two- site probabilities.
With this approximation, the 
two-dimensional PA dynamics are written as follows:
\begin{eqnarray}
a_2(n+1) &=& p_2^2 P_n(\bullet\bullet\bullet) +
p_1 p_2 \{P_n(\bullet\bullet\circ)+P_n(\circ\bullet\bullet) \}\nonumber\\
&+& p_1^2 \{ P_n(\bullet\circ\bullet)+P_n(\circ\bullet\circ)  \}
\nonumber\\
&\cong& \frac{(2 p_2 a_2(n)+p_1 a_1(n))^2}{4b_1(n)} +
\frac{p_1^2 a_1(n)^2}{4b_0(n)},\label{eq:pair1}\\
a_1(n+1) &\cong& p_2(1-p_2) \frac{2a_2(n)^2}{b_1(n)} \nonumber\\
&+& (p_1+p_2-2 p_1 p_2) \frac{a_1(n)a_2(n)}{b_1(n)}\nonumber\\
&+& p_1 (1-p_1) \frac{a_1(n)^2}{2 b_0(n)b_1(n)}
+ p_1 \frac{a_1(n)a_0(n)}{b_0(n)},\label{eq:pair2}
\end{eqnarray} 
where $b_1(n) = a_2(n) + a_1(n)/2$, $b_0(n) = 1 - b_1(n)$ and
$a_0(n) = 1 - a_1(n) - a_2(n)$. Trajectories of the PA
dynamics are shown in \FIG\ref{fig:pair}(a) for two sets of
supercritical parameter sets: $(p_1, p_2)=$ (0.52, 1) (thin
lines) and (0.66,
0.66) (thick lines). 
In accordance with \FIG\ref{fig:dk}(a), the individual PA
dynamics own stable fixed points near (0, 1) and (0.308, 0.145).
However, as shown in \FIG\ref{fig:pair}(b), the population dies out
when they alternate.
Although the supercritical parameter region of the PA dynamics
deviates from that of the DK counterparts, the results for
the PA dynamics qualitatively agree with those for the DK dynamics
shown in \FIG\ref{fig:dk}.

The Parrondo's paradox is unlikely to happen in
one-dimensional systems since they lack auxiliary dimensions
that counteract the 
seeming tendency of population increase. To demonstrate this,
let us imagine the simplistic mean-field approximation in which a joint
probability is approximated by a product of single-site
probabilities (e.g. $P_n(\circ\bullet)\cong P_n(\circ)P_n(\bullet)$). The
approximate one-dimensional system is written as
\begin{equation}
b_1(n+1)= p_2
b_1(n)^2 + 2p_1 b_1(n)(1-b_1(n)),
\end{equation}
which has fixed points $b_1 = 0$ and
$b_1 = (2p_1-1)/(2p_1-p_2)$ \cite{Gutowitz,Tome,Harada,Atman02}. Let us
pick two mean-field
dynamical systems so that their nontrivial fixed points
are positive and stable, with $2p_1>p_2$
and $p_1>1/2$ satisfied.  Then,
when two mean-field dynamics alternate,
the particle density $b_1(n)$ just moves
between these two fixed points in the long run.
Accordingly, the population never dies out, and no paradoxical
phenomenon occurs.

\section{Canonical Model}

To generalize the Parrondo's paradox found for the DK and PA dynamics, we
construct a simple canonical model with dimension two, which is
the presumed minimal degree of freedom
for the paradox. As we have mentioned, the relevant
features of the DK and PA dynamics can be summarized as follows.

\begin{description}

\item[(i)] Trajectories of dynamics $A$ and those of dynamics $B$
transverse in the way 
as shown in \FIGS\ref{fig:dk}(a) and \ref{fig:pair}(a).  More
specifically, in the $a_1$-$a_2$ space,
the slope of a trajectory of dynamics $A$ (thin lines)
is less negative than that of a trajectory
of dynamics $B$ (thick lines) at the crossing point, at least in a
certain region.

\item[(ii)] Each of $A$ and $B$ is not applied
too many times successively. In other words,
$k$ in the sequence $A^k B^k$, $A^k B$, or $A B^k$ should be small enough,
as explained with \FIGS\ref{fig:dk}(d) and \ref{fig:sdk}(a).

\item[(iii)] Population change
rates are proportional to the population size. To weaken the condition
may result in the same conclusion just with a different
convergence rate. Here we assume this linearity for our canonical model.

\end{description}

The paradox also relies on the 
following implicit assumptions.

\begin{description}

\item[(iv)] Dynamics $A$ and dynamics $B$ have
sufficiently separated nontrivial fixed points.

\item[(v)] The origin is
the deterministically unstable but stochastically reachable
fixed point for both $A$ and $B$. 

\end{description}

Figure~\ref{fig:dk}(a) and \ref{fig:pair}(a) further indicate that
the origin and the nontrivial fixed point are connected by
applying $B$ infinitely many times (thick lines).
However, it is not true for $A$ (thin lines)
because $A$ is nongeneric in the
sense that all the points on the $a_2$ axis are fixed
points. Actually, no point with $a_1=0$ and $0<a_2<1$
 is realizable because it would mean that 
two consecutive sites take state $\bullet\bullet$ and
$\circ\circ$ with positive probabilities
but not $\bullet\circ$ or $\circ\bullet$.
The total size of the boundaries between
clusters of $\bullet$ and those of $\circ$, or $a_1(n)$,
determines the population change
rates \cite{Durrettbook}. It declines to zero as a point in the
$a_1$-$a_2$ space approaches the $a_2$ axis. 

\begin{figure}[h]
\begin{center}
\includegraphics[height=2.75in,width=2.75in]{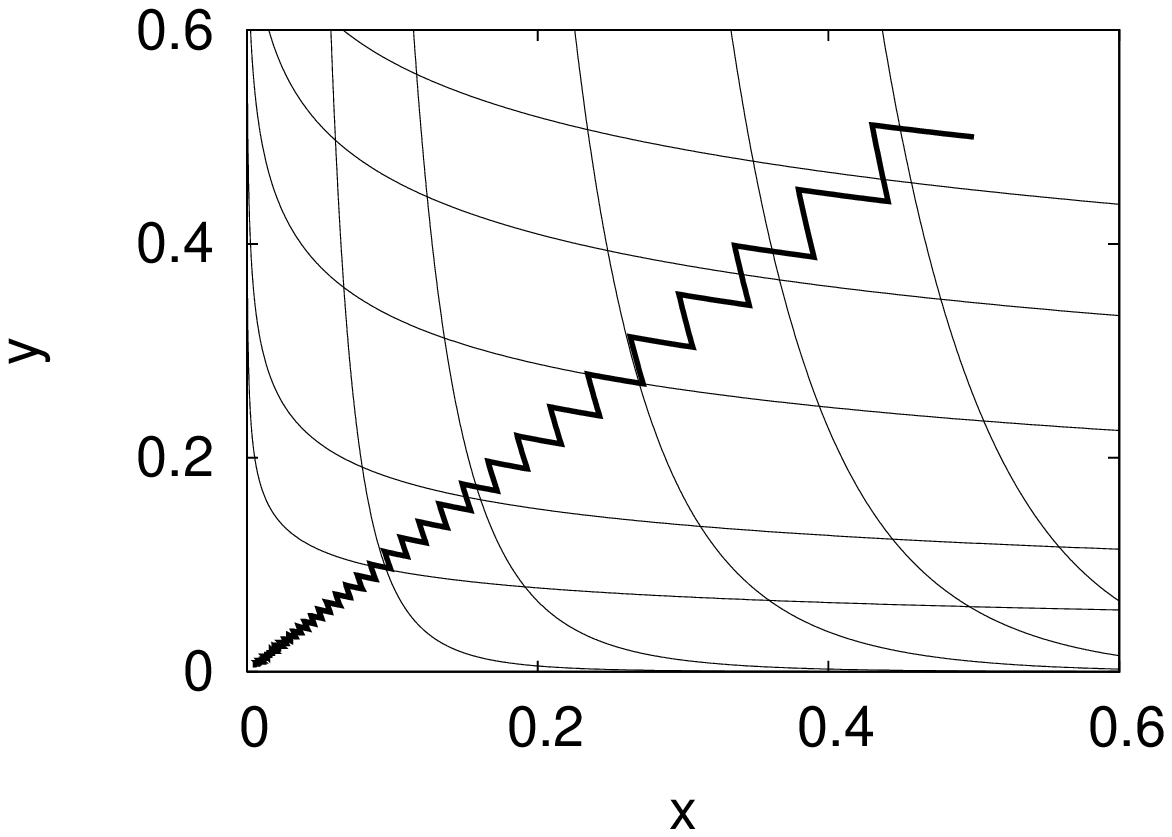}
\caption{}
\label{fig:canonical}
\end{center}
\end{figure}

In fact, we have chosen ($p_1,
p_2$)=(0.52, 1) for dynamics $A$ 
just because the obtained DK dynamics is rigorously
supercritical. The paradoxical dynamics appear robustly
against changes in $p_1$ and $p_2$, which can make the dynamics
generic. With this in mind,
we construct a two-dimensional continuous-time system
that satisfies the conditions listed above.
We propose to alternate
two dynamical systems:\\
{\bf dynamics $A$}
\begin{eqnarray*}
\dot{x}&=&-x,\\
\dot{y}&=&\lambda y(1-y),
\end{eqnarray*}
and\\
{\bf dynamics $B$}
\begin{eqnarray*}
\dot{x}&=&\lambda x(1-x),\\
\dot{y}&=&-y,
\end{eqnarray*}
where $0<x, y<1$.
The properties (iii), (iv), and (v) are obviously
satisfied, with (iii) also supported by \FIG\ref{fig:stat} (squares). Both
dynamics $A$ and $B$
have a saddle at the origin. The point ($0, 1$) of $A$
and ($1, 0$) of $B$ are stable equilibria,
and each of them is connected to the origin by
a heteroclinic orbit. The property (i) is satisfied if $0<\lambda<1$.
To guarantee (ii), we set $\lambda=0.3$
and the duration of each dynamics equal to 0.15.
Figure~\ref{fig:canonical} summarizes flows of the individual dynamics
(thin lines) and those of the alternating dynamics (thick lines).
We again observe the paradox that
the alternating dynamics lead the state toward the origin.

\section{Conclusions}

We have shown using the DK model and its simplifications
that mixtures of two supercritical dynamics can yield
subcritical dynamics in which the population dies out.  This
counterintuitive behavior
occurs if individual component dynamics have at least dimension two
and satisfy certain criteria.
The property (i) is characteristic of the DK or the
canonical model, and it agrees with 
some natural occasions but not with others
\cite{Murraybook}. The other
four requirements do not seem to spoil 
the reality. The properties (iii) and (v) are
satisfied when production rates are primarily proportional to the
population mass, which is quite common for ecological and social
systems \cite{Murraybook}. Periodical and random environmental changes comply
with (ii) and (iv). Such changes
may be also caused by continual, periodic, or random human control of a system
with the aim of moving the stable fixed point to more desirable one.
However, our results indicate that environmental changes or oddly
managed control measures can cause a total disaster even if the
system instantaneously stays in a supercritical `good' regime all the time.
The other way round, there is a general expect that a situation that is
subcritical at any moment can be changed into a supercritical one with
appropriate controls, which is originally illuminated by 
the Parrondo's paradox
\cite{Harmer,Parrondo,Harmer02,Amengual}. In the context of the
Parrondo's paradox, our model provides another
mechanism of its occurrence in addition to inhomogeneous game rules or
players with memory \cite{Parrondo}, namely, spatial extension.

\begin{figure}[h]
\begin{center}
\includegraphics[height=1.4in,width=2.75in]{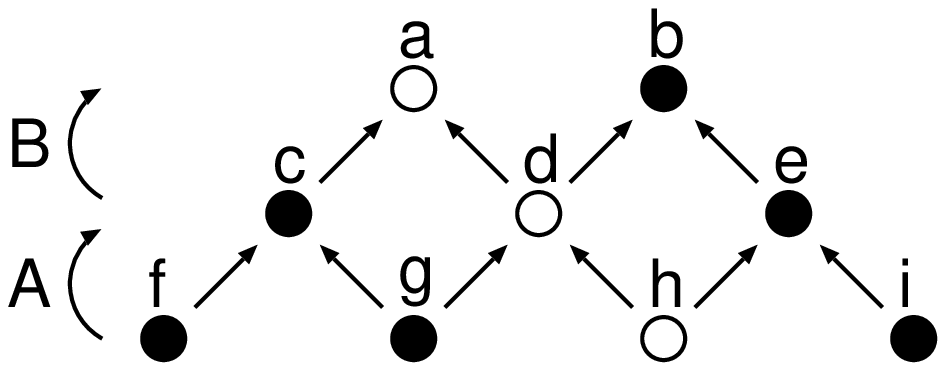}
\caption{}
\label{fig:long}
\end{center}
\end{figure}

Lastly, we can regard a block of sequence of $A$ and $B$, such as
$A^k B$ and $AB^k$, as a transformation done in just one step. By
doing so, the alternating DK model seems similar to $m$-neighborhood
probabilistic cellular automata (PCA) with $m\ge 3$.  For PCA, phase
diagrams have been studied in simple cases where the dynamical rule
depends only on the number of particles in the neighborhood with $m=3$
\cite{Bagnoli,Atman03}.  However, the model proposed here is more
complex even with the simplest sequence $ABAB\ldots$, which should be
compared to PCA with $m=3$.  One reason is that outcomes depend not
only on the number but also on the arrangements of particles in a
neighborhood \cite{Martins}. For instance, it is easy to verify
$P_n(\bullet|\bullet\circ\bullet)$ $\neq$
$P_n(\bullet|\bullet\bullet\circ)$. More importantly, the alternating
DK dynamics are not special cases of finite-range PCA. To illustrate
this, let us consider $ABAB\ldots$. In
\FIG\ref{fig:long}, the state of site $a$, which we write $\xi(a)$
depends on $\xi(f)$, $\xi(g)$, and $\xi(h)$, while $\xi(g)$, $\xi(h)$,
and $\xi(i)$ put together determine $\xi(b)$.  In 3-neighborhood
PCA, there exists no intermediate layer of sites such as $c$, $d$, and
$e$. Therefore, once $\xi(f)$, $\xi(g)$, $\xi(h)$, and $\xi(i)$ are
given, $\xi(a)$ and $\xi(b)$ are independent.  On
the other hand, in our model,
$\xi(a)$ and $\xi(b)$ are partially correlated, or 
correlated even conditioned by
$\xi(f)$, $\xi(g)$, $\xi(h)$, and $\xi(i)$. This is because
both $\xi(a)$ and $\xi(b)$ depend on
$\xi(d)$. By the same token, the infinite-range correlation
is generated just after single application of $AB$, which prohibits
use of powerful duality equations
\cite{Konno02a,Konno02b,Katori04}. In this sense, our model
stiupulates a class of infinite particle systems different
from ordinary PCA.  However, on the analogy of the
Parrondo's paradox, the phenomena reported in this paper may hold for
PCA and more general alternating dynamics with general
neighborhood sizes.

\begin{acknowledgement}
We thank K. Sato for his helpful comments.
This study is supported by the Grant-in-Aid for Scientific Research
(JSPS Fellows) and the Grant-in-Aid for Scientific Research (B)
(No.12440024) of Japan Society of the Promotion of Science.
\end{acknowledgement}

\newpage

Figure captions

\bigskip

Figure 1: Schematic diagram showing the DK probabilistic cellular automaton.

\bigskip

Figure 2: (a) Trajectories of the DK model for 
dynamics A
with $(p_1, p_2) =$ ($0.52, 1$) (thin lines) and those for
dynamics B with $(p_1,
p_2) =$ ($0.76, 0.76$) (thick lines). The other panels show 
population dynamics when
we repeat (b) $AB$, (c) $A^4 B$, and (d) $A^{30}B$.
The initial conditions for (b, c, d) are $(a_1(0),
a_2(0))=(0.5, 0.25)$.

\bigskip

Figure 3: Dynamics of the population size when we repeat
(a) $A^k B^k$ with $k=1$ (thinnest line), $2$, $3$, and
$4$ (thickest line), (b)
$A^k B$ with $k=1$ (thinnest), $2$, $4$,
$15$, $30$ (thickest), and (c) $A B^k$ with
$k=1$ (thinnest), $2$, $3$ and $4$ (thickest). In (b), the lowermost 
line corresponds to $k=4$. In (c), 
the upper lines, which are nearly superimposed,
 correspond to $k=1$ and $3$, whereas the lower
lines correspond to $k=2$ and $4$.

\bigskip

Figure 4: Population dynamics when
$A$ and $B$ randomly appear with probability $r$ and $1-r$,
respectively. (a) $r=0.02$, $0.05$, $0.1$, $0.2$ (from upper to lower
lines),
and (b) $r=0.2$, $0.4$, $0.5$, $0.75$ (from lower to upper lines).

\bigskip

Figure 5: Population change rates in terms of the population size
for the DK (crosses), PA (circles), and canonical (squares) dynamics.
The change rates for the DK and PA dynamics
are measured by the Euclidean distances of two points with unit time
difference in the
$a_1$-$a_2$ space.  The population size is equal to
$a_1(t)/2+a_2(t)$ for the DK and PA dynamics and is defined to be
$\sqrt{x^2+y^2}$ for the canonical dynamics.

\bigskip

Figure 6: (a) Trajectories of the PA dynamics for $(p_1,p_2)=$
($0.52, 1$) (thin lines) and those for $(p_1, p_2) =$ ($0.66,
0.66$) (thick lines). (b) Population dynamics in the alternating PA
dynamics starting from $(a_1(0), a_2(0)) = (0.5, 0.25)$.

\bigskip

Figure 7: Trajectories of dynamics $A$ and those of dynamics $B$
of
the canonical model (thin lines), superimposed by those of the alternating
dynamics starting from $(x,y)=(0.5, 0.5)$ (thick line).

Figure 8: Alternating dynamics with sequence $AB$
compared with standard 3-neighborhood PCA.

\clearpage

\clearpage

\onecolumn

\clearpage

\clearpage

\clearpage

\clearpage

\clearpage

\clearpage

\begin{thebibliography}{Vanvreeswijk99}

\bibitem{Wolframbook}
S. Wolfram, {\it Theory and Applications of Cellular Automata}
(World Scientific, Singapore, 1986)

\bibitem{Domany}
E. Domany, W. Kinzel,
% Equivalence of cellular automata to Ising models
% and directed percolation.
Phys. Rev. Lett. {\bf 53}, 311 (1984)
%(4), --314

\bibitem{Kinzel}
W. Kinzel,
% Phase transitions of cellular automata. 
Z. Phys. B Condens. Matter {\bf 58}, 229 (1985)
%--244

\bibitem{Harmer}
G. P. Harmer, D. Abbott, Nature {\bf 402}, 846 (1999)
% Losing strategies can win by Parrondo's paradox.

\bibitem{Parrondo}
J. M. R. Parrondo, G. P. Harmer, D. Abbott,
% New paradoxical games based on Brownian ratchets.
Phys. Rev. Lett. {\bf 85}, 5226 (2000)
%(24), --5229

\bibitem{Harmer02}
G. P. Harmer, D. Abbott, 
% A review of Parondo's paradox
Fluctuation and Noise Letters {\bf 2}, R71 (2002).
%(2) --R107

\bibitem{Durrettbook}
R. Durrett, {\it Lecture Notes on Particle Systems
and Percolation} (Wadsworth, Inc., California, 1988)

\bibitem{Martins} 
M. L. Martins, H. F. Verona de Resende, C. Tsallis,
A. C. N. Magalh\~{a}es,
% Evidence for a new phase in the Domany-Kinzel Cellular Automaton.
Phys. Rev. Lett. {\bf 66}, 2045
(1991)
%  (15), --2047

\bibitem{Gutowitz}
H. A. Gutowitz, J. D. Victor, B. W. Knight,
% Local structure theory for cellular automata.
Physica D {\bf 28}, 18 (1987)
%--48

\bibitem{Tome}
T. Tom\'{e},
% Spreading of damage in the Domany-Kinzel
% cellular automaton: a mean-field approach
Physica A {\bf 212}, 99 (1994)
%--109

\bibitem{Harada}
Y. Harada, H. Ezoe, Y. Iwasa, H. Matsuda, K. Sato,
% Population persistence and spatially limited social interaction.
Theor. Popul. Biol. {\bf 48}, 65 (1995)
%--91

\bibitem{Atman02}
A. P. F. Atman, R. Dickman,
%Quasistationary distributions for the Domany-Kinzel stochastic cellular automaton. 
Phys. Rev. E {\bf 66}, 046135 (2002)

\bibitem{Bagnoli}
F. Bagnoli, N. Boccara, R. Rechtman,
% Nature of phase transitions in a probabilistic cellular automaton
% with two absorbing states.
Phys. Rev. E {\bf 63}, 046116 (2001)

\bibitem{Atman03}
A. P. F. Atman, R. Dickman, J. G. Moreira,
% Phase diagram of a probabilistic cellular automaton with three-site
% interactions.
Phys. Rev. E {\bf 67}, 016107 (2003)

\bibitem{Katori}
M. Katori, N. Konno, H. Tanemura,
% Survival probabilities for discrete-time models in one dimension.
J. Stat. Phys. {\bf 99}, 603 (2000)
%No.1/2, --612

\bibitem{Konno02a}
N. Konno, 
%Dualities for a class of finite-range probabilistic cellular automata 
% in one dimension.
J. Stat. Phys. {\bf 106}, 915 (2002)
%(516), --922

\bibitem{Konno02b}
N. Konno,
%Self-duality for multi-state probabilistic cellular automata
% with finite range interactions.
J. Stat. Phys.
{\bf 106} 923 (2002)
%(516), --930.

\bibitem{Katori04}
M. Katori, N. Konno, A. Sudbury, H. Tanemura.
% Dualities for the Domany-Kinzel model.
J. Theo. Prob.
{\bf 17} 131 (2004)
%(1) --144

\bibitem{Liggett95}
T. M. Liggett,
% Survival of discrete time growth models, with applications to oriented percolation. 
Ann. Applied. Prob. {\bf 5}, 613 (1995)
%--636

\bibitem{Onody}
R. N. Onody, U. P. C. Neves,
%Series expansion of the directed percolation probability.
J. Phys. A: Math. Gen. {\bf 25}, 6609 (1992)
%--6615 

\bibitem{Jensen}
I. Jensen, A. J. Guttmann, 
% Series expansions of the percolation probability for directed
% square and honeycomb lattices.
J. Phys. A: Math. Gen. {\bf 28}, 4813 (1995)
%--4833

\bibitem{Amengual}
P. Amengual, A. Allison, R. Toral, D. Abbott
% Discrete-time ratchets, the Fokker-Planck equation and Parrondo's paradox. 
Proc. R. Soc. Lond. A {\bf 460}, 2269 (2004).

\bibitem{Murraybook}
J. D. Murray, {\it Mathematical Biology, I: An Introduction, Third
Edition} (Springer-Verlag, New York, 2002)

\end{thebibliography}
\end{document}